\documentstyle[aps,prl,preprint,floats,epsfig]{revtex}

\begin{document}

\preprint{\tighten\vbox{\hbox{\hfil CLEO CONF 00-6}
                        \hbox{\hfil ICHEP 00-552}
}} 

\title{Study of
         {\boldmath  $\chi_{c1}$} and   {\boldmath  $\chi_{c2}$} meson
         production in {\boldmath $B$} meson decays}  

\author{CLEO Collaboration}
\date{\today}

\maketitle
\tighten

\begin{abstract} 
Using    a   sample   of
$9.7\times10^6$ $B  \overline B$ meson pairs collected with the CLEO
detector,    we study inclusive $B$
 meson decays  to  the $\chi_{c1}$  and $\chi_{c2}$  charmonia  states.
We measure the  branching fraction for the inclusive $\chi_{c1}$ production 
in  $B$  decays to be 
${\cal B}(B\to\chi_{c1}X)=(4.14\pm0.31\pm 0.40)\times10^{-3}$,
where the  first  uncertainty is statistical  and  the  second one  is
 systematic. 
 We  obtain the branching fractions  for direct $\chi_{c1}$ and
$\chi_{c2}$  production in  $B$  decays by 
subtracting   the contribution  from the decay chain 
$B\to\psi(2S)X$ with $\psi(2S)\to  \chi_{c1,2} \gamma$.  We measure ${\cal
B}(B\to\chi_{c1}[{\rm direct}]X)=(3.83\pm0.31\pm 0.40)\times10^{-3}$.
  No  statistically   significant signal   for  $\chi_{c2}$
production is observed in either case. We determine the 95\% C.L. upper limits to be ${\cal B}(B\to\chi_{c2}X)<2.0\times10^{-3}$,  ${\cal B}(B\to\chi_{c2}[{\rm direct}]X)<1.7\times10^{-3}$, and 
 ${\cal B}(B\to\chi_{c2}[{\rm direct}]X)/{\cal B}(B\to\chi_{c1}[{\rm direct}]X)<0.44$. All quoted results are preliminary.
\end{abstract}
\vfill
\begin{flushleft}
.\dotfill .
\end{flushleft}
\begin{center}
Submitted to XXXth International Conference on High Energy Physics, July
2000, Osaka, Japan
\end{center}

\newpage

{
\renewcommand{\thefootnote}{\fnsymbol{footnote}}

\begin{center}
G.~Brandenburg,$^{1}$ A.~Ershov,$^{1}$ Y.~S.~Gao,$^{1}$
D.~Y.-J.~Kim,$^{1}$ R.~Wilson,$^{1}$
T.~E.~Browder,$^{2}$ Y.~Li,$^{2}$ J.~L.~Rodriguez,$^{2}$
H.~Yamamoto,$^{2}$
T.~Bergfeld,$^{3}$
% ...
\end{center}
 
\small
\begin{center}
$^{1}${Harvard University, Cambridge, Massachusetts 02138}\\
$^{2}${University of Hawaii at Manoa, Honolulu, Hawaii 96822}\\
$^{3}${University of Illinois, Urbana-Champaign, Illinois 61801}\\
% ...
\end{center}
\setcounter{footnote}{0}
}
\newpage

The recent measurements  of charmonium production in
various high-energy physics reactions  have brought welcome  surprises
and challenged  our  understanding both of   heavy-quark production
and of  quarkonium  bound  state formation.
 The  measurement  of a  
large  production rate of high-$P_T$ charmonium   at     the 
Tevatron~\cite{CDF-D0-charmonia} was in sharp disagreement with the 
then-standard color-singlet model.
The development of the  NRQCD factorization framework~\cite{Bodwin:1995jh} 
has put the calculations  of the inclusive charmonium 
production    on a rigorous  footing.
The high-$P_T$ charmonium
production rate at the Tevatron is now well understood in this formalism. 
The recent CDF 
measurement of  charmonium polarization~\cite{CDF-polarization}, 
however, appears to disagree with the NRQCD prediction.
The older color-evaporation model accommodates both the high-$P_T$
charmonium production rate and polarization measurements at the 
Tevatron~\cite{Amundson:1997qr}. 

Inclusive $B$ meson decays to  charmonia 
is another area to confront theoretical predictions with experimental  data.
The color-singlet contribution, for example, is a factor of 5--10
below~\cite{Beneke:1999ks}   the observed inclusive $J/\psi$ production 
rate~\cite{Balest:1995jf}.
A  measurement of the 
$\chi_{c2}$-to-$\chi_{c1}$ production ratio in $B$ decays provides 
an  especially clean  test of the charmonium production models. 
       The        $V-A$         current        
 $\overline  c \gamma_{\mu}(1-\gamma_{5})c$ cannot create 
a  $c \overline c$ pair in a 
${^{2S+1}L}_J = {^3P}_2$ state,   therefore  the  decay
$B\to   \chi_{c2}  X$  is  forbidden     
 at   leading  order in $\alpha_s$ in the color-singlet model~\cite{chi-color-singlet}.
The importance of the 
color-octet mechanism for the $\chi_{c}$ production in $B$ decays  
was recognized~\cite{Bodwin:1992qr} even before the development of the 
NRQCD framework. The NRQCD calculations
of the $B$ decays to charmonia at  the next-to-leading order in $\alpha_s$ 
can be found in Ref.~\cite{Beneke:1999ks}.   
The color-octet contribution to 
 $B\to\chi_{cJ}X$ decays is proportional to $2J+1$. Therefore, 
if this mechanism dominates, then the $\chi_{c2}$-to-$\chi_{c1}$
production ratio should be 5:3. On the contrary, if the  color-singlet contribution dominates, then the ratio should be 0:1.
The color-evaporation model  predicts $\chi_{c2}:\chi_{c1}=5:3$~\cite{Schuler:1999az}.

Our data  were collected at   the Cornell Electron Storage Ring
(CESR) with    two   configurations    of     the   CLEO detector
called CLEO~II~\cite{Kubota:1992ww}  and
CLEO~II.V~\cite{Hill:1998ea}.  The components of the CLEO detector
most relevant to this analysis are the charged particle tracking
system, the CsI electromagnetic calorimeter, and the muon chambers.
In CLEO~II the momenta of charged particles are measured in a tracking
system consisting of a 6-layer straw tube chamber,  a 10-layer
precision drift chamber, and a 51-layer main drift chamber, all
operating inside a 1.5 T  solenoidal magnet. The main drift chamber
also provides a measurement of the  specific ionization, $dE/dx$, used
for particle identification.  For  CLEO~II.V, the straw tube  chamber
was replaced  with a  3-layer silicon vertex detector, and the gas in
the main drift chamber was changed from an argon-ethane to a
helium-propane mixture. The muon chambers  consist of proportional
counters placed at increasing depth in  the steel absorber. 
 
 We use
9.2~$\rm fb^{-1}$ of $e^+e^-$ data taken at the $\Upsilon(4S)$
resonance    and 4.6~$\rm fb^{-1}$ taken  
60~MeV below the $\Upsilon(4S)$ resonance (off-$\Upsilon(4S)$ sample).  
Two thirds of the data  were collected with
the CLEO~II.V detector.  The simulated  event samples used in this
analysis were generated with a GEANT-based~\cite{GEANT} simulation of
the  CLEO detector response and  were processed in a similar manner as
the  data.  

We reconstruct the $\chi_{c1}$ and $\chi_{c2}$ radiative decays to $J/\psi$. 
The branching fractions for 
the $\chi_{c1,2}\to J/\psi \, \gamma$ decays  are, respectively,   
$(27.3\pm1.6)\%$ and $(13.5\pm1.1)\%$~\cite{PDG}, whereas the branching fraction  for the $\chi_{c0}\to J/\psi \, \gamma$ decay is only $(0.66\pm0.18)\%$.
In addition, the $\chi_{c0}$ production rate in $B$ decays is expected 
to be smaller  than the $\chi_{c1,2}$ rates~\cite{Beneke:1999ks,Bodwin:1992qr}.
 We therefore do not attempt to measure  
$\chi_{c0}$ production  in this analysis. 

\begin{figure}[htb]
\centering 
\epsfxsize=14cm 
\epsfbox{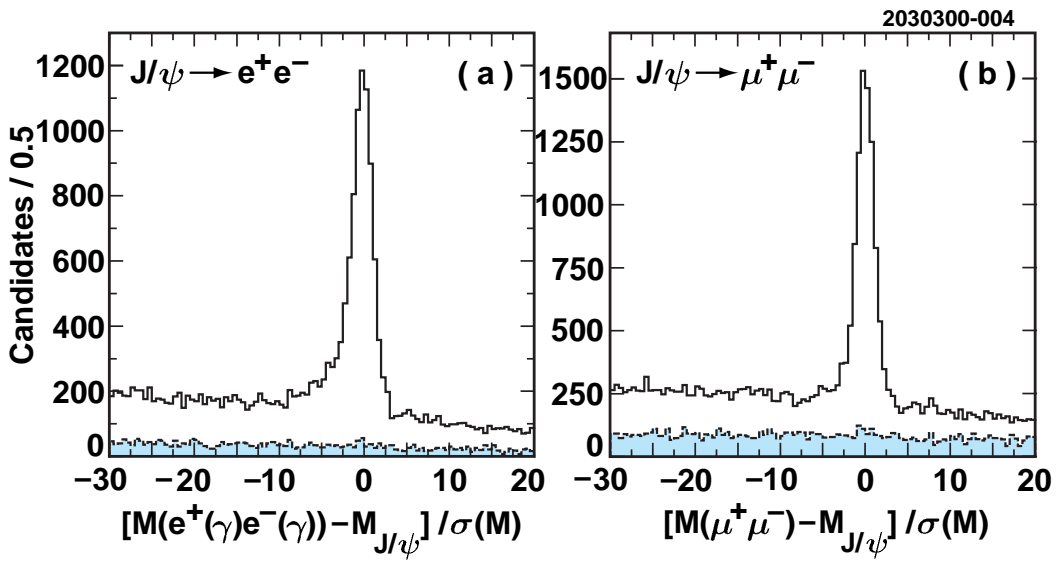}
\caption{Normalized invariant mass of the (a) $J/\psi \to e^+ e^-$ and
(b) $J/\psi  \to \mu^+  \mu^-$ candidates  in  the data. The momentum of
the  $J/\psi$ candidates is required to be less than 2~GeV/$c$.    
The  shaded  histogram  represents the
luminosity-scaled  data taken  60~MeV below the $\Upsilon(4S)$
showing  the level of background from non-$B \overline B$ events.  } 
\label{fig:data_ee_mumu_on_off}
\end{figure}

 We reconstruct  both  $J/\psi \to e^+ e^-$  and  $J/\psi \to \mu^+
\mu^-$  decays.  Electron candidates are
identified based on the ratio of the  track momentum to the associated
shower energy in the CsI calorimeter and on  the $dE/dx$  measurement.
The internal bremsstrahlung  in the  $J/\psi  \to  e^+ e^-$ decay  as
well  as the bremsstrahlung in the detector  material produces a long
radiative tail in the  $ e^+ e^-$   invariant mass distribution  and
impedes efficient $J/\psi  \to   e^+  e^-$  detection.   We  recover
some of the  bremsstrahlung photons by selecting the photon  shower
with the smallest opening angle with respect to the  direction of the
$e^\pm$ track evaluated at the interaction point, and then   requiring
this opening angle to be smaller than $5^\circ$. We therefore refer to
the $e^+ (\gamma) e^- (\gamma)$ invariant mass when we describe the
$J/\psi \to e^+ e^-$ reconstruction.  For the $J/\psi \to \mu^+ \mu^-$
reconstruction, one of the muon candidates is  required to penetrate
the steel absorber to a depth greater than 3  nuclear interaction
lengths.  We relax the absorber penetration  requirement for the
second muon candidate  if it is not expected to reach a muon chamber
either because its energy is too low or because it does not point to
a region of the  detector  covered by the muon  chambers.  For these
muon candidates  we require the ionization  signature in the  CsI
calorimeter to be  consistent with that of a muon.
We use the normalized invariant mass for the $J/\psi$ candidate selection.
For example, the
normalized  $J/\psi \to \mu^+ \mu^-$ mass is defined as  $[M(\mu^+
\mu^-)-M_{J/\psi}]/\sigma(M)$, where $M_{J/\psi}$ is the world average
value of the $J/\psi$ mass~\cite{PDG} and $\sigma(M)$ is the
calculated mass resolution for that  particular  $\mu^+ \mu^-$
combination. The average $\ell^+ \ell^-$ invariant mass resolution  is
12~MeV$/c^2$.  The normalized mass distributions for the $J/\psi \to
\ell^+ \ell^-$     candidates   are         shown     in
Fig.~\ref{fig:data_ee_mumu_on_off}.   We   require     the normalized
mass  to be from $-6$  to $+3$  for the $J/\psi  \to  e^+  e^-$ and
from $-4$  to $+3$ for the  $J/\psi \to  \mu^+   \mu^-$ candidates.
The momentum of
the  $J/\psi$ candidates is required to be less than 2~GeV/$c$, which
is slightly above  the maximal $J/\psi$ momentum in  $B$ decays. 

 Photon candidates   for $\chi_{c1,2}\to J/\psi \, \gamma$  reconstruction 
must be detected in the barrel region of the calorimeter, 
defined as $|\cos\theta_{\gamma}|<0.71$,
where $\theta_{\gamma}$ is the angle between the beam axis and the
candidate photon.  
Most of the  photons in  $\Upsilon(4S) \to B \overline  B$
events   come from $\pi^0$ decays.  We therefore do not use a photon for
the $\chi_{c1,2}\to J/\psi \, \gamma$ reconstruction 
if it  can be paired with another photon to produce a $\pi^0$
candidate with the  normalized $\pi^0 \to \gamma \gamma$  mass between
$-3$ and $+2$. 

\begin{figure}[htb]
\centering                                            
\epsfxsize=7.0cm
\epsfbox{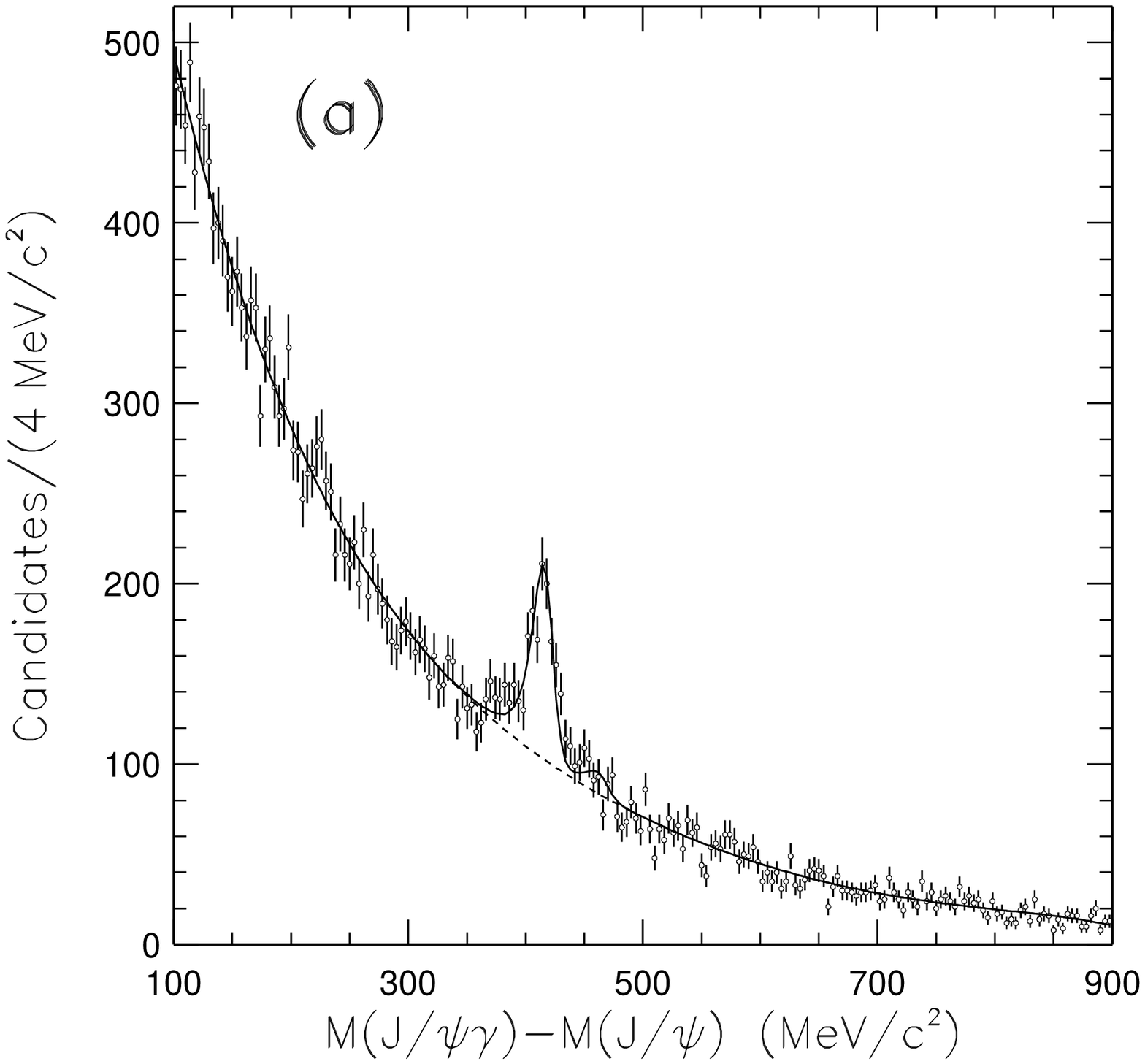}
\epsfxsize=7.0cm
\epsfbox{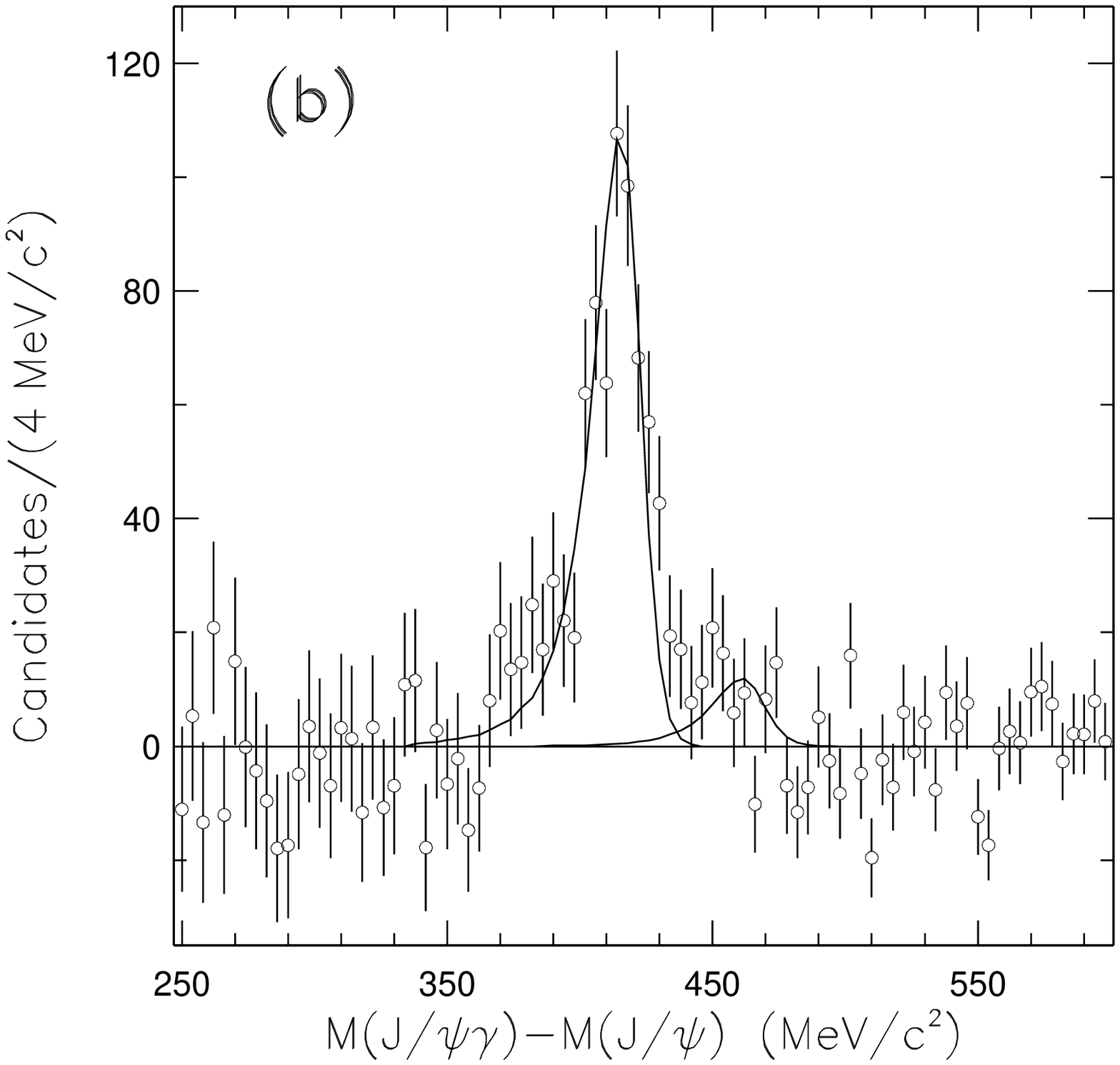}
\caption{Plot (a) shows the  $M(J/\psi \gamma)-M(J/\psi)$ distribution in the $\Upsilon(4S)$ data (points with error bars).
The fit function is shown by a solid line with the background component represented by a dashed line. Plot (b) shows the background-subtracted distribution in the 
signal region  with 
the $\chi_{c1}$ and $\chi_{c2}$ fit components  represented by a solid line.}
\label{fig:chi_incl_data_fit}
\end{figure}

We determine the $\chi_{c1}$   and  $\chi_{c2}$  yields in 
a maximum-likelihood fit to the mass-difference distribution $M(J/\psi \gamma)-M(J/\psi)$ (Fig.~\ref{fig:chi_incl_data_fit}), 
where $M(J/\psi)$ is the measured  mass of a $J/\psi$ candidate. 
The excellent electromagnetic calorimeter allows us to resolve the 
$\chi_{c1}$   and  $\chi_{c2}$ peaks. The $M(J/\psi \gamma)-M(J/\psi)$ mass-difference resolution is approximately 8~MeV/$c^2$ and is dominated by the photon energy resolution. The background in the fit is  approximated with a 
5th order Chebyshev polynomial; all the polynomial coefficients are allowed to float.
The $\chi_{c1}$   and  $\chi_{c2}$ signal shapes are fit with 
 templates extracted from Monte Carlo simulation;  only  the template 
normalizations are floating in the fit.
The    $\chi_{c1}$   and  $\chi_{c2}$   signal   yields in  the $\Upsilon(4S)$ data
  are  $N^{\rm ON}(\chi_{c1})=672\pm47[{\rm stat}]$  and $N^{\rm ON}(\chi_{c2})=83\pm37[{\rm stat}]$.
The  $\chi_{c1}$   and  $\chi_{c2}$  yields in off-$\Upsilon(4S)$ data 
are   consistent with zero:     
$N^{\rm  OFF}(\chi_{c1})=4.1\pm7.1[{\rm stat}]$   
and   $N^{\rm OFF}(\chi_{c2})=1.1\pm6.5[{\rm stat}]$. 
Subtracting the contributions from non-$B\overline B$ continuum 
events, we obtain the total inclusive $B\to \chi_{c1}X $   and  $B\to\chi_{c2}X$  yields 
\begin{displaymath}
N(B\to \chi_{c1}X)=664\pm49[{\rm stat}]\;{\rm and}\;N(B\to \chi_{c2}X)=81\pm39[{\rm stat}].
\end{displaymath}

Taking into account the systematic uncertainties associated with the fit, 
we determine  the $B\to \chi_{c2}X$ signal yield significance to be  $2.0$ standard deviations. Subtracting the feeddown from the decay chain 
 $B\to \psi(2S)X$ with $\psi(2S)\to\chi_{c2}\gamma$ and accounting for the 
associated systematic uncertainty,  
we likewise determine the  
 significance of the evidence for the direct  $B\to \chi_{c2}X$ decays 
to be  $1.4$ standard deviations. 

To calculate branching fractions for the $B\to \chi_{c1,2}X$ decays,
we use the measured 
signal yields $N(B\to \chi_{c1,2}X)$, reconstruction efficiencies 
extracted from simulation, the number of produced $B
\overline B$  pairs, and the daughter branching fractions.
For the calculation of the rates  for direct $B\to \chi_{c1,2}X$ decays, we make an assumption that the only other source of  $\chi_{c1,2}$ meson 
production  in 
$B$ decays is the decay chain  $B\to \psi(2S)X$ with $\psi(2S)\to\chi_{c1,2}\gamma$.  
The resulting branching fractions are listed in 
Table~\ref{tab:results}.
\begin{table}[htb]
\center
\caption{\small The measured branching fractions for inclusive 
$B$ decays to $\chi_{c1}$ and $\chi_{c2}$. We subtract  the 
$\psi(2S)\to \chi_{c1,2} \gamma$ feeddown to arrive at the  branching 
fractions for  direct  $B\to\chi_{c1,2}X$ decays. The correlation between   
the uncertainties is taken into  account in the calculation of the 
$\chi_{c2}$-to-$\chi_{c1}$ production ratio (last row).}
\label{tab:results}
\begin{tabular}{lll} 
Branching fraction   &  Measured value    & 
95\% C.L.  \\ 
or ratio  &    &  upper limit~{\protect \cite{Feldman:1998qc}}  \\ \hline
${\cal B}(B\to\chi_{c1}X)$ & $(4.14\pm0.31\pm 0.40)\times10^{-3}$ & --- \\ 
~~${\cal B}(B\to\chi_{c1}[{\rm direct}]X)$ & $(3.83\pm0.31\pm0.40)\times10^{-3}$ & --- \\ \hline
${\cal B}(B\to\chi_{c2}X)$ & $(0.98\pm0.48\pm 0.15)\times10^{-3}$ & $<2.0\times10^{-3}$~\tablenotemark[1] \\
~~${\cal B}(B\to\chi_{c2}[{\rm direct}]X)$ & $(0.71\pm0.48\pm 0.16)\times10^{-3}$ & $<1.7\times10^{-3}$ \\ \hline
{\large $\frac{{\cal B}(B\to\chi_{c2}[{\rm direct}]X)}{{\cal B}(B\to\chi_{c1}[{\rm direct}]X)}$} & $0.18\pm0.13\pm 0.04$ & $<0.44$ \\ 
\end{tabular}
\flushleft \tablenotetext[1]{The  95\% confidence interval in the ``unified'' approach ~{\protect \cite{Feldman:1998qc}} is $[0.2; 2.0]\times10^{-3}$.}
\end{table}

The systematic uncertainties are listed in Table~\ref{tab:systematics}. 
The sources of the uncertainty  can be grouped into three categories:

(i) {\it Fit procedure.---}
This category includes the systematic uncertainties  due to the choice of 
 the signal and  background shapes as well as the   bin size.
To fit the $\chi_{c1}$   and  $\chi_{c2}$ signal, we use 
the templates extracted from simulation.  We therefore are sensitive  to 
the imperfections in the  simulation of the photon  energy measurement.   The systematic uncertainties associated with the simulation of 
the calorimeter 
response are estimated by comparing the $\pi^0\to \gamma\gamma$ invariant mass lineshapes for inclusive 
$\pi^0$ candidates in data and  Monte Carlo  samples. Then we modify the 
$\chi_{c1}$   and  $\chi_{c2}$ templates to determine the resulting uncertainty in the signal yields.
To estimate the uncertainty associated with the calorimeter  energy scale, 
we   shift  the  $\chi_{c1}$   and  $\chi_{c2}$ templates 
by $\pm0.6$~MeV/$c^2$ 
in the fit. The uncertainty due to  time-dependent  variations of the 
calorimeter  energy scale is small compared to the 
overall energy scale uncertainty.
To estimate the uncertainty due the calorimeter  energy resolution,
we change the width of the $\chi_{c1}$   and  $\chi_{c2}$ templates by 
$\pm4\%$. The uncertainty in background shape is probed by fitting the background with the  template extracted from the high-statistics samples of 
simulated $\Upsilon(4S) \to B \overline  B$ and non-$B \overline  B$
continuum events; only the template normalization, not its shape, is allowed to float in the fit. 

(ii) {\it Efficiency calculation.---}
This category includes the   uncertainties in   the number of produced $B
\overline B$  pairs,   tracking efficiency, photon   detection efficiency,  lepton  detection efficiency, 
statistical uncertainties  of the simulated  event  samples, and the model-dependence in  the simulation of $B\to\chi_{c1,2}X$ decays. 
The angular distribution of the  $\chi_{c1,2}\to J/\psi \gamma$ decays
affects the   photon energy spectrum. 
 We  define  the helicity angle
$\theta_h$  to  be  the  angle  between  the  $\gamma$  direction   in
$\chi_{c}$ frame and the  $\chi_{c}$  direction in  the $B$ frame.  We
assume flat $\cos\theta_h$  distribution in our simulation of $B\to\chi_{c1,2}X$ decays. Systematic uncertainty associated with this assumption is estimated by comparing the reconstruction efficiencies in the Monte Carlo  samples with $I(\theta_h)\propto \sin^2\theta_h$  and 
$I(\theta_h)\propto \cos^2\theta_h$ angular distributions.
Another source of uncertainty is the modeling of the  $X$ system 
in the simulation of $B\to\chi_{c1,2}X$ decays. The
 photon detection efficiency depends on the  assumed model  
through  the $\chi_c$ momentum spectrum and  the $\pi^0$ multiplicity of  the final state. In our 
simulation, we assume that $X$ is either $K$ or   one of the 
higher $K$ resonances:
24\%  $K$,  24\%  $K^*(892)$, 14\% $K_1(1270)$,   
14\%    $K_1(1400)$,  13\%   $K^*_0(1430)$,    and 11\% $K^*_2(1430)$; we also include the decay chain  $B    \to  \psi(2S)  X$ with $\psi(2S)\to\chi_{c1,2}\gamma$. To estimate the systematic uncertainty, we compare the $\chi_{c}\to J/\psi \gamma$ detection efficiency extracted  using this sample with the efficiency in the sample where we assume that $X$ is either a  $K^+$ or $K^0_S\to\pi^+\pi^-$.

(iii) {\it  Assumed branching fractions.---}
This category includes the   uncertainties on  the  branching fractions.   
 We use the following values of the daughter  branching fractions:
${\cal  B}(J/\psi \to \ell^+  \ell^-)=(5.894\pm0.086)\%$~\cite{Bai:1998di},
 ${\cal B}(\chi_{c1} \to     J/\psi \gamma)=(27.3\pm1.6)\%$~\cite{PDG}, 
and  ${\cal B}(\chi_{c2} \to J/\psi \gamma)=(13.5\pm1.1)\%$~\cite{PDG}.
In calculation of ${\cal B}(B\to\chi_{c1,2}[{\rm direct}]X)$ 
we also  assume the following values: ${\cal B}(B\to \psi(2S) X)= (3.5\pm0.5)\times10^{-3}$~\cite{PDG},  ${\cal B}(\psi(2S) \to \chi_{c1} \gamma)=(8.7\pm0.8)\%$~\cite{PDG}, and ${\cal B}(\psi(2S)\to\chi_{c2} \gamma)=(7.8\pm0.8)\%$~\cite{PDG}. 
\begin{table}[htb]
\center
\caption{\small Systematic uncertainties on the measured branching fractions.}
\label{tab:systematics}
\begin{tabular}{l|c|c} 
Source of & \multicolumn{2}{c}{relative uncertainty in \%} \\ 
systematic  uncertainty & ${\cal B}(B\to\chi_{c1}X)$ & 
${\cal B}(B\to\chi_{c2}X)$  \\ \hline
(i)~~~Fit procedure   &  &  \\ 
~~~~~~~Calorimeter energy scale & $0.4$ & $5.6$ \\ 
~~~~~~~Calorimeter energy  resolution & $2.8$ & $6.9$  \\ 
~~~~~~~Background shape & $1.8$ & $6.8$ \\
~~~~~~~Bin size & $0.0$ & $1.9$ \\ 
(ii)~~Efficiency calculation   &  &  \\ 
~~~~~~~$N(B \overline B)$ & $2.0$ & $2.0$ \\ 
~~~~~~~Tracking efficiency & $2.0$ & $2.0$ \\ 
~~~~~~~Lepton identification & $4.2$ & $4.2$ \\ 
~~~~~~~Photon finding & $2.5$ & $2.5$ \\ 
~~~~~~~Monte Carlo statistics & $0.7$ & $0.7$ \\ 
~~~~~~~Composition of $X$ in $B\to\chi_{c1,2}X$ simulation & $3.3$ & $3.3$  \\
~~~~~~~Angular distribution for $\chi_{c1,2}\to J/\psi \gamma$ & $1.0$
& $1.0$ \\ 
(iii)~Assumed branching fractions   &  &  \\ 
~~~~~~~${\cal B}(\chi_{c1,2}\to J/\psi \gamma)$ & $5.9$ & $8.1$ \\ 
~~~~~~~${\cal B}(J/\psi \to \ell^+ \ell^-)$ & $1.5$ & $1.5$ \\ 
~~~~~~~${\cal B}(B\to\psi(2S)X)$\tablenotemark[1] & $1.1$ & $5.5$ \\ 
~~~~~~~${\cal B}(\psi(2S)\to\chi_{c1,2}\gamma)$\tablenotemark[1]  & $0.7$ & $4.0$ \\ 
\end{tabular}
\flushleft \tablenotetext[1]{Contributes only to uncertainty on ${\cal B}(B\to\chi_{c1,2}[{\rm direct}]X)$.}
\end{table}

In conclusion, we have measured the branching fractions for  inclusive 
$B$ meson decays   to   $\chi_{c1}$  and $\chi_{c2}$  charmonia  states.
Our measurements are consistent with and improve upon the previous 
CLEO results~\cite{Balest:1995jf}. Our measurement of the branching ratio for
direct $\chi_{c2}$ and $\chi_{c1}$ production in $B$ decays
is consistent with the prediction of the color-singlet model~\cite{chi-color-singlet} and disagrees with the color-evaporation model~\cite{Schuler:1999az}.
In NRQCD framework, our measurement suggests that the color-octet 
mechanism does not dominate in $B\to\chi_{c}X$ decays.
           
We   gratefully acknowledge the effort of the CESR staff in providing us
with excellent luminosity and running conditions.  This work was
supported by  the National Science Foundation, the U.S. Department of
Energy, the Research Corporation, the Natural Sciences and Engineering
Research Council of Canada,  the A.P. Sloan Foundation,  the Swiss
National Science Foundation,  the Texas Advanced Research Program, and
the Alexander von Humboldt Stiftung.


\begin{thebibliography}{99}

\bibitem{CDF-D0-charmonia} CDF Collaboration, F.~Abe {\it et al.},
Phys.\ Rev.\ Lett.\ {\bf 69}, 3704 (1992); {\bf 79}, 572 (1997); {\bf 79}, 578 (1997); D0
Collaboration, S.~Abachi {\it et al.}, Phys.\ Lett.\ B {\bf 370}, 239
(1996).

\bibitem{Bodwin:1995jh} G.~T.~Bodwin, E.~Braaten and G.~P.~Lepage,
Phys.\ Rev.\ D {\bf 51}, 1125 (1995).


\bibitem{CDF-polarization} CDF Collaboration, T.~Affolder {\it et
al.}, Report No. FERMILAB-PUB-00-090-E, hep-ex/0004027 (submitted to
Phys.~Rev.~Lett.).

\bibitem{Amundson:1997qr} J.~F.~Amundson, O.~J.~Eboli, E.~M.~Gregores, 
and F.~Halzen, Phys.\ Lett.\ B {\bf 390}, 323 (1997)



\bibitem{Beneke:1999ks} M.~Beneke, F.~Maltoni, and I.~Z.~Rothstein,
Phys.\ Rev.\ D {\bf 59}, 054003 (1999).


\bibitem{Balest:1995jf}
CLEO Collaboration, R.~Balest {\it et al.},
Phys.\ Rev.\  D {\bf 52}, 2661 (1995).

\bibitem{chi-color-singlet} J.~H.~Kuhn, S.~Nussinov, and R.~Ruckl, Z.\
Phys.\ {\bf C5}, 117 (1980); J.~H.~Kuhn and R.~Ruckl, Phys.\ Lett.\
{\bf 135B}, 477 (1984); Phys.\ Lett.\ B {\bf 258}, 499 (1991).


\bibitem{Bodwin:1992qr} G.~T.~Bodwin, E.~Braaten, T.~C.~Yuan, and
G.~P.~Lepage, Phys.\ Rev.\ D {\bf 46}, 3703 (1992).

\bibitem{Schuler:1999az} G.~A.~Schuler, Eur.\ Phys.\ J.\ C {\bf 8}, 273
(1999).


\bibitem{Kubota:1992ww} CLEO Collaboration, Y.~Kubota {\it et al.},
Nucl.\ Instrum.\ Meth.\ Phys.\ Res.\ A {\bf 320}, 66 (1992).


\bibitem{Hill:1998ea} T.S.~Hill, Nucl.\ Instrum.\ Meth.\ Phys.\ Res.\
A {\bf 418}, 32 (1998).

\bibitem{GEANT} CERN Program Library Long Writeup W5013 (1993).


\bibitem{PDG} Particle Data Group, C.~Caso {\it et al.},   Eur.\
Phys.\ J.\ C {\bf 3}, 1 (1998).

\bibitem{Bai:1998di} BES Collaboration, J.~Z.~Bai {\it et al.}, Phys.\
Rev.\  D {\bf 58}, 092006 (1998).

\bibitem{Feldman:1998qc} G.J.~Feldman and R.D.~Cousins,  Phys.\ Rev.\
D {\bf 57}, 3873 (1998). 
\end{thebibliography}
\end{document}